REST: Routing with En-route Seating Thresholds for pedestrian accessibility

Nir FULMAN[1a,b], Himshikhar Mazumdar[a,b]

[a] *Institute of Geography, GIScience Chair, Heidelberg University, Heidelberg, Germany*
[b] *HeiGIT at Heidelberg University, Heidelberg, Germany*

Abstract

Public seating can support walking by providing opportunities to rest at useful positions along complete journeys. We introduce a route-based framework that assesses nearby seating, access to everyday destinations, and whether rests are well spaced along routes. One method finds the shortest route that does not exceed a chosen distance without rest; another allows longer stretches but adds a cost expressed as equivalent walking distance. We apply the framework to Mannheim, Germany. With a 1,000 m travel budget and a 500 m limit on walking without an opportunity to rest, 99.5% of older residents in Feudenheim lived within 500 m of a rest, yet only 33.5% could reach at least one destination in each of five categories—food shops, pharmacies, primary healthcare, public transport and community services—on ordinary shortest routes meeting this limit. Finding alternative routes with sufficiently spaced rests increased this share to 61.4%; allowing longer stretches with an added cost raised it to 76.2%, compared with 89.7% when resting needs were ignored. Across Mannheim, 20.2% of older residents lost access to a reachable category when routes required a resting opportunity every 500 m, versus 7.1% when longer stretches were allowed at extra cost. Compared with the shortest physical distance, the least burdensome routes incurred extra costs from seating gaps: a median of 7.3 m of equivalent walking distance across city areas (90th percentile: 45.9 m). Better-placed seating could reduce this burden without changing street layouts. These measures help planners identify deficiencies, test improvements and guide route choices.

Highlights

- A unified framework makes journey-level resting accessibility measurable and routable.
- Strict routing finds the shortest journey that satisfies a maximum no-rest interval.
- Soft routing retains threshold-exceeding trips and measures their excess burden.
- Strict continuity affected 20.2% of older residents; 7.1% remained unresolved under soft routing.
- The framework extends to tourism, charging, refuelling and scheduled driver rests.

[1] Corresponding Author: Nir Fulman, nir.fulman@uni-heidelberg.de.

## 1. Introduction

Regular walking supports physical health, helps prevent or manage chronic conditions, and contributes to psychological well-being and cognitive functioning in later life (Saelens and Handy, 2008; Shah et al., 2026). As a mode of everyday transport, it connects older adults to shops, healthcare, public transport, social contacts and other destinations that sustain independent living (Shigematsu et al., 2009; World Health Organization Regional Office for Europe, 2017). When functional limitations make walking difficult, older adults tend to be less physically active even where neighbourhood conditions would otherwise support walking (Portegijs et al., 2017; Tuomola et al., 2024). The resulting loss of mobility can also narrow access to services, social interaction and life outside the home, compromising autonomy and participation (World Health Organization, 2015). These consequences concern a growing share of the world's population: 830 million people, or 10.2% of the global population, were aged 65 years or older in 2024, and this number is projected to reach 1.7 billion, or 17.3%, by 2054 (United Nations Department of Economic and Social Affairs, Population Division, 2024). Supporting walking in later life is therefore a matter of public health, transport accessibility and social inclusion.

The built environment helps determine whether walking is a practical and attractive option in later life. At the neighbourhood scale, the distribution of destinations and the connectivity of the pedestrian network determine whether everyday activities can be reached by reasonably direct routes; systematic evidence associates such conditions with older adults' active travel (Cerin et al., 2017). Along those routes, continuous and well-maintained pavements, manageable gradients, safe crossings and separation from traffic influence the effort and risk involved in walking (Moran et al., 2017; Fulman et al., 2025a). Shade, greenery, lighting, visual quality and low noise further affect thermal comfort, perceived safety and the experience of the journey (Van Cauwenberg et al., 2018). These attributes can facilitate walking by reducing physical or psychological demands, or discourage it by making an otherwise short journey uncomfortable, unsafe or difficult to negotiate. Walking accessibility therefore depends not only on the distance between homes and destinations, but also on the environmental conditions encountered between them (Kan et al., 2026).

Public seating supports walking by giving pedestrians an opportunity to rest before continuing a journey. This function is particularly important in later life, when limited endurance and other functional limitations can make walking without an opportunity to rest difficult (Simonsick et al., 2008; Vestergaard et al., 2009). Walk-along interviews in Flanders identified public seating among the facilities relevant to trips to nearby destinations (Van Cauwenberg et al., 2012). Bench-focused interviews in Vancouver described seating as a mobility aid that supported use of outdoor space and social contact (Ottoni et al., 2016). A photographic experiment also found that streets containing a bench were rated as more inviting for transport walking (Van Cauwenberg et al., 2014). Together, these studies identify public seating as part of the infrastructure supporting older adults' walking and participation in public life.

The need for opportunities to rest is not uniform across older adults because ageing does not produce a single level of walking capacity. Functional limitations also interact with perceived neighbourhood walkability (Suzuki et al., 2021). Differences in endurance, pain, balance, sensory function, chronic conditions and reliance on mobility aids can alter how demanding walking without an opportunity to rest is (World Health Organization, 2015; Fulman et al., 2025a). Evidence on seating preferences reflects this variation. In a survey of 1,761 Norwegian adults aged 67 years and older, the share rating additional benches beside pavements as very important to walking more rose from 17% among those aged 67-69 to 46% among those aged 85-92 (Krogstad et al., 2015). A photographic choice experiment with 1,131 Flemish older adults identified a subgroup, representing about 7% of the sample, for whom the presence of a bench and vegetation were the most influential street features; this subgroup also contained the highest proportion of walking-aid users (Van Cauwenberg et al., 2016). These findings show that sensitivity to seating varies, but they do not quantify how opportunities to rest influence actual walking behaviour.

To assess how resting needs affect access to everyday destinations, one approach is to treat the capacity of walking without an opportunity to rest as a hard limit. UK mobility guidance recommends distance limits without rest of 50 m for people using a walking stick, 100 m for mobility-impaired people without an aid, and 150 m for wheelchair users and visually impaired people, while emphasizing substantial variation between individuals and conditions (UK Department for Transport, 2022). Bench-planning research similarly evaluates provision against maximum intervals between resting opportunities, commonly between 50 and 150 m (Usui and Hino, 2018; Ptak-Wojciechowska et al., 2025). Such intervals can be understood as precautionary provision standards: they seek to ensure that a resting opportunity is available before an individual's walking capacity is exhausted, rather than implying that every user will stop at each opportunity. Under this interpretation, a journey meets the selected threshold only when every stretch from the origin to the first rest, between successive rests and from the final rest to the destination is no longer than that threshold. A bench near one end of the journey cannot compensate for a longer gap elsewhere. This produces a strict rest-continuity criterion: exceeding the threshold makes the route non-compliant for the specified capability or design scenario.

Resting needs can, however, affect how demanding a journey becomes before they prevent its completion. Research on independently mobile older adults distinguishes the effort experienced during walking from the ability to complete a walking task (Simonsick et al., 2014). Valiani et al. (2016), for example, recorded steadily increasing perceived effort as older adults with chronic fatigue progressed through a 400 m walk. Huber et al. (2026) examined how walking without an opportunity to rest relates to everyday route choice, using 972 errand choice sets reported by 573 older adults in Mannheim, Uppsala and Kayseri and comparing the chosen routes with 3,629 plausible alternatives between the same origins and destinations. Bench locations were ordered along each route, and the study measured the proportion of route length contained in the part of the longest no-bench interval beyond 500 m. Routes with more walking beyond this threshold were less likely to be chosen; a ten-percentage-point increase in the measure had the same estimated utility loss as adding 31 m of walking in Mannheim, 49 m in Uppsala and 91 m in Kayseri. Exceeding 500 m therefore reduced route preference rather than making a route unavailable in the model. Comparable distances also appear in older adults' stated walking preferences: an Australian survey found that 500–1,000 m was the most frequently reported preferred distance for reaching destinations, including among respondents aged 80 years and older (Garrard, 2013). Together, these studies support a complementary soft interpretation in which excessive walking without rest remains possible but carries an additional cost, with resting opportunities ending the bouts over which that cost accumulates.

Applying either interpretation citywide requires connecting resting opportunities to the journeys people need to make. Existing bench assessments take two broad approaches. Some describe infrastructure supply through counts and densities (Akinci et al., 2021; Urra-Uriarte et al., 2024). These measures show where seating is abundant or scarce, but a high density of benches does not establish whether they provide continuity along a journey. Other approaches measure the gaps between resting opportunities. Usui and Hino (2018) derived the distribution of continuous walking distances from nearest-neighbour distances and applied the method around Tokyo Station. Usui (2024) used the distance from a bench to its furthest neighbouring bench and tested where additional benches would most reduce the probability of gaps exceeding 100 m. Ptak-Wojciechowska et al. (2025) assigned benches to streets, compared their supply with the number implied by a selected interval, classified the streets and simulated new locations. Koch and Legeby (2022) used the pedestrian network to identify connected street segments on which benches occur within specified intervals. These methods reveal shortages and discontinuities that counts and densities can conceal. However, a gap on one street may be avoided through an alternative route, while adequately spaced benches elsewhere may not help someone reach a particular destination. Gap assessments therefore do not directly establish whether a suitable route exists between an origin and destination, how much additional walking it requires, or how demanding the journey remains.

Pedestrian routing provides a way to assess these journey-level consequences. Routing algorithms calculate paths through connected street networks, commonly identifying the shortest distance or travel time between an origin and a destination. Beyond guiding individual journeys, these calculations support accessibility analysis by determining which destinations can be reached from residential locations and the travel required to reach them (Geurs and van Wee, 2004; Fulman et al., 2025a). Repeating these calculations across a city reveals spatial differences in access to everyday destinations. However, shortest-distance routes do not account for environmental conditions that may make walking more or less suitable. Quality-aware routing systems address this by trading additional distance against greenery, shade, quietness, pollution and other environmental attributes (Novack et al., 2018; Ludwig et al., 2021; Helle et al., 2023; Feng et al., 2024; Siriaraya et al., 2020). Age-friendly route planners have also begun to include benches and other supportive infrastructure. Aranguren et al. (2024), for example, assign nearby benches, toilets and drinking fountains to street segments and allow users to favour routes containing more of these amenities. This brings resting infrastructure into route calculation, but the amenities are counted rather than evaluated by their ordered positions. Such an approach works for general amenity provision, but it cannot represent the specific function of resting opportunities. Two benches clustered at one location do not provide twice the continuity, and a bench near the destination cannot relieve a long no-rest stretch earlier in the journey. The cost of the next metre depends on how far the person has already walked since the previous rest, and encountering a bench resets that accumulated distance. A strict formulation must reject a route when this distance exceeds the selected limit, whereas a soft formulation must increase its cost without removing it from consideration. Both therefore require routing that tracks distance since the previous rest, rather than assigning each street segment a fixed cost independent of the path taken to reach it.

Representing either formulation requires a shift in the logic of the routing algorithm. Standard shortest-path methods such as Dijkstra's algorithm efficiently identify a least-cost path by retaining the cheapest arrival at each intersection. This is valid with fixed additive costs because a costlier arrival cannot become cheaper after following the same continuation (Dijkstra, 1959). The strict formulation has a general precedent in shortest-path problems with replenishment: a resource accumulates along the path, specified locations reset it, and paths exceeding its limit are infeasible (Smith et al., 2012). Here, distance walked since the previous rest is the resource and benches are the replenishment opportunities. In both rest formulations, arrivals at the same intersection can differ in accumulated cost and distance since the previous rest. A slightly costlier but more recently rested arrival may allow a continuation that would be infeasible or more costly from a cheaper arrival. The search must therefore retain alternatives that could still produce a better complete route. Established label-based shortest-path methods provide this mechanism by storing alternative arrivals as separate records, called labels, and discarding only those that cannot improve the final solution (Martins, 1984; Irnich and Desaulniers, 2005; Maristany de las Casas et al., 2021).

We introduce two complementary journey-level formulations of resting accessibility. The strict rest-continuity formulation finds the shortest route on which no stretch of walking without an opportunity to rest exceeds the selected distance limit. The soft rest-sensitive formulation instead finds the route with the lowest generalized walking cost, combining physical distance with an additional penalty for walking beyond the threshold without rest. We first demonstrate the different analytical uses of the framework in Feudenheim, a neighbourhood of Mannheim, using building-level maps. We then aggregate the same origin-level route outcomes across Mannheim to assess citywide heterogeneity.

## 2. Methods

Our analysis combines a pedestrian network, mapped resting opportunities, population-weighted residential origins and everyday destinations in Mannheim. We examine seating supply and proximity before comparing ordinary shortest routes with two rest-sensitive alternatives (Table 1).

Table 1. Analytical stages and questions in the Mannheim study.

| Stage | Calculation | Question answered |
|---|---|---|
| Seating supply and proximity | Counts and densities; nearest-rest distance along the network and from residential origins. | Where is mapped seating located, and how close is a network location or home to a first reachable rest? |
| Shortest-path access and threshold compliance | Shortest route within the walking budget; Longest stretch without an opportunity to rest on that same route assessed against the selected distance threshold. | What does pedestrian accessibility indicate, and would its ordinary shortest route already meet the selected threshold? |
| Strict rest-continuity routing | Shortest route within the walking budget on which no stretch without an opportunity to rest exceeds the selected distance limit; additional physical distance relative to the shortest route. | What access remains under a maximum-gap requirement, and how much additional walking is needed to recover a non-compliant shortest-route connection? |
| Soft rest-sensitive routing | Lowest generalized walking cost, combining physical distance with a penalty proportional to walking without rest beyond the resting threshold; excess walking burden relative to shortest distance. | Which journeys remain within a distance-equivalent budget when a long gap is burdensome rather than prohibited, and how much more burdensome are they than the physical shortest path? |
| Reporting scale | Building- and street-level results in Feudenheim; 500 m hexagonal aggregation of the same outcomes across Mannheim. | How do local route mechanisms translate into heterogeneous citywide planning conditions? |

## 2.1. Study area

The study area is the administrative territory of Mannheim, a regional centre in southwestern Germany. Mannheim had 329,998 residents 2025. Of these, 62,266 (18.9%) were aged 65 years or older (Stadt Mannheim, 2026a). In 2025, the share aged at least 80 years exceeded 10% in Sonnenschein, Vogelstang and Niederfeld, but was below 3% in Jungbusch, Luzenberg and Neckarstadt-West (Stadt Mannheim, 2026a). Mannheim ranges from the compact orthogonal grid of the Quadrate to former villages, housing estates, industrial areas and peripheral landscapes (Stadt Mannheim, 2025). Feudenheim, a 6.07 $km^2$ neighbourhood in the north-eastern part of Mannheim, is used as the local case suitable for local planning.

### 2.2. Spatial data and preparation

#### 2.2.1. Pedestrian network and resting opportunities

Residential origins and population weights are restricted to Mannheim's municipal territory. The pedestrian network, resting opportunities, and everyday destinations, are extracted for Mannheim and a 2,000 m buffer around it. This buffer exceeds the maximum route length within the study budget (1,000 m), with additional allowance for network matching distances. Origins outside Mannheim are excluded. OpenStreetMap supplies the directed pedestrian network and resting-opportunity inventory, prepared with OSMnx (Boeing, 2017; OpenStreetMap contributors, 2026).

The rest inventory combines objects tagged amenity=bench, public-transport stops, platforms or shelters explicitly tagged bench=yes, and objects tagged leisure=picnic_table or picnic_table=yes. Features tagged as private, inaccessible, customer-only or permit-only are excluded. Rest features are converted to points and matched to the nearest physical pedestrian edge within 20 m. Matched points are stored as metric offsets in both permitted traversal directions. Rest positions coinciding within 0.1 m on the same physical edge are treated as one resting opportunity. The suitability of the OSM pedestrian network and resting-opportunity data was assessed through comparison with official ATKIS data and targeted street-level imagery checks, as detailed in Appendix B.1 and B.2.

#### 2.2.2. Residential origins and older-population weights

The German 2022 Census 100 m grids provide total population and the share aged 65 years or older (Statistische Aemter des Bundes und der Laender, 2024). For 554 of 3,872 cells with a missing older-adult share, the citywide observed share of 19.24% is used. To represent where older residents live within each cell, we allocate its estimated older population to residential buildings, using 27,589 OpenStreetMap residential-building footprints within Mannheim's municipal boundary. Each building receives a share proportional to its residential capacity, calculated as footprint area multiplied by building levels. Recorded OpenStreetMap building:levels values are used where available; buildings without a usable value are assigned one level. The influence of this assumption on population-weighted accessibility results was examined in a sensitivity check (Appendix B.3).

The Census allocation initially estimates 60,325.8 residents aged at least 65 years and is proportionally rescaled citywide to the 2025 municipal total of 62,266. Cells without a receiving residential building remain as grid-centroid origins. Residential-building and centroid origins were snapped to their nearest pedestrian-network node within 75 m. Of 766 residual Census-cell centroids created for populated cells without a receiving residential building, 714 matched the pedestrian network within 75 m and were retained. Together with 26,707 matched residential-building points, they were consolidated to 10,313 weighted network origin nodes. Source points representing about 899 estimated older residents lay beyond 75 m of the network and were excluded before the matched weights were rescaled to the municipal total.

### 2.2.3. Everyday destinations

Everyday destinations are grouped into five recurrent-need categories (Table 2). The five categories were selected to cover distinct everyday needs relevant to older adults' walking and independent living (Cerin et al., 2017). Destinations are matched to the nearest pedestrian-network node within 75 m.

Table 2. Destination categories and OpenStreetMap definitions.

| Destination category | Included OpenStreetMap tags |
|---|---|
| Food shops | shop=supermarket, convenience, bakery, butcher, greengrocer or deli |
| Pharmacies | amenity=pharmacy or healthcare=pharmacy |
| Primary healthcare | amenity=doctors or clinic; healthcare=doctor, general_practitioner or clinic |
| Public transport | public_transport=platform; highway=bus_stop; railway=tram_stop, halt or station |
| Community services | amenity=community_centre, social_facility or library |

## 2.3. Seating supply and proximity

We count resting opportunities and calculate three density measures: rests per square kilometer, per kilometer of physical pedestrian network and per 1,000 residents aged at least 65 years. Each pedestrian-network segment is counted once even when the routing graph contains one directed edge for each travel direction. These calculations are restricted to the administrative boundary.

To measure proximity to resting opportunities, we divide each pedestrian-network segment into sections no longer than 25 m and evaluate the midpoint of each section. For every midpoint, we calculate the shortest walking distance along the network to the nearest rest. Midpoints are restricted to Mannheim, but the nearest rest may be outside the city. Each midpoint is weighted by the length of its section. We report the percentiles and the share of the network within 100, 250 and 500 m of a rest. We repeat the calculation at residential origins, weighted by the number of older residents.

## 2.4. Route models and rest-sensitive routing

### 2.4.1. Definitions and resting-distance measures

Throughout the route models, $o$ and $d$ denote an origin and destination, $P$ a candidate route, and $L(P)$ its physical length. Each mapped resting opportunity is assumed to provide a usable rest that ends the current bout without additional cost. The parameters $T$ and $B$ are the reference distance for walking without an opportunity to rest and the total journey budget, respectively; $\lambda$ gives the added distance-equivalent cost for each metre walked beyond $T$ without rest.

For a route $P$ of physical length $L(P)$, the ordered rest positions divide the route into without-rest walking bouts $r_1, \ldots, r_m$. The origin begins the first bout and the destination ends the last. Two quantities describe its resting conditions. The maximum no-rest gap is

$$G(P) = \max_i r_i \quad (1)$$

For a no-rest distance $T$, the total excess no-rest distance is

$$E_T(P) = \sum_i \max(0, r_i - T) \quad (2)$$

2.4.2. Route objectives and constraints

The baseline shortest route between origin $o$ and destination $d$ minimizes physical distance. Its cost is

$$D_{\text{od}} = \min_P L(P) \quad (3)$$

This condition provides the reference against which rest-related detours and costs are calculated.

The strict rest-continuity formulation sets a maximum distance that can be walked without an opportunity to rest and applies this limit to the entire journey. It identifies the shortest route satisfying the selected threshold:

$$H_{\text{od}}(T) = \min_{P:G(P)\leq T} L(P) \quad (4)$$

If no compliant route exists, $H_{\text{od}}(T)$ is infinite.

The soft model assigns route $P$ the generalized walking cost

$$C_{T,\lambda}(P) = L(P) + \lambda E_T(P) \quad (5)$$

The minimum soft cost is

$$C_{\text{od}}(T, \lambda) = \min_P C_{T,\lambda}(P) \quad (6)$$

2.4.3. Routing algorithms and common search procedure

The baseline is computed with standard Dijkstra shortest-path search on the same directed pedestrian network (Dijkstra, 1959).

Both rest-sensitive formulations are shortest-path problems where the cost or feasibility of the next step depends not only on the street segment itself, but also on how far the pedestrian has already walked since the most recent rest. In the strict formulation, distance since rest is a resource that may not exceed $T$; this is a rest-specific instance of a shortest-path problem with resource replenishment (Smith et al., 2012). In the soft formulation, exceeding the threshold remains permissible but increases the cost of subsequent walking. Both algorithms use a multi-label search, which can retain several alternative arrivals at the same network location as separate records, or labels (Martins, 1984; Irnich and Desaulniers, 2005).

Each label records the current network node and the street segment used to reach it; together, these identify its traversal state. The incoming segment is recorded because the algorithm prohibits immediately turning back along that same segment. The label also stores accumulated cost, distance since the most recent rest, and a link to the preceding label used to reconstruct the route.

Within the pedestrian network, resting opportunities located on an edge are processed in travel order by dividing the edge into subsegments at their offsets. The length of each subsegment is charged before the distance since rest is reset at its ending rest. Origins and destinations are represented by their matched network nodes.

At a given traversal state, a label is dominated if another has no greater accumulated cost and no greater distance since rest. The algorithm retains the remaining, nondominated labels and places new labels in a priority queue, which selects the lowest-cost label first. A label that becomes dominated is marked inactive and skipped if it is later removed from the queue.

2.4.4. Strict rest-continuity search algorithm

For the strict search, accumulated physical distance is $d$, and $s$ is distance since the most recent rest. Extending a label across a subsegment of length $x$ produces $d' = d + x$ and $s' = s + x$. The candidate is infeasible and is discarded if $s' > T$. If the subsegment ends at a rest, the feasibility check is made first and $s$' is then reset to zero.

At the same traversal state, strict label $A$ dominates strict label $B$ when

$$d_A \leq d_B, s_A \leq s_B \quad (7)$$

The first active destination label returned gives the shortest route satisfying the resting limit. If no such label is found, no compliant route exists.

Algorithm 1. Strict rest-continuity shortest-path search.

```
INPUT  graph G, origin o, destination d, threshold T
Q ← priority queue ordered by physical distance
A[q] ← active labels at traversal state q
insert Label(o, distance = 0, sinceRest = 0) into Q and A[o]

while Q is not empty do
   L ← extract the label with minimum distance
   if L is inactive then continue
   if L is at d then return reconstructPath(L)

   for each permitted outgoing edge e do
      C ← copy(L)
      for each ordered subsegment x of e do
         C.distance ← C.distance + x
         C.sinceRest ← C.sinceRest + x
         if C.sinceRest > T then
            discard C; stop traversing e
         if x ends at a rest then C.sinceRest ← 0
      if C was discarded then continue
      if some label in A[state(C)] dominates C then discard C
      else
         mark every label dominated by C inactive
         insert C into A[state(C)] and Q

return NO COMPLIANT PATH
```

Appendix A.2 establishes the correctness and termination of Algorithm 1.

2.4.5. Soft rest-sensitive search algorithm

For the soft search, accumulated generalized cost is $g$ and threshold violations remain feasible. For a bench-free subsegment of length $x$ entered with distance since rest $s$, the additional excess distance created by the subsegment is

$$\Delta E(s, x; T) = \max(0, s + x - T) - \max(0, s - T) \quad (8)$$

The transition produces $g' = g + x + \lambda \, \Delta \, E$ and $s' = s + x$. If the subsegment ends at a rest, $s$' is reset to zero after its cost has been charged.

At the same traversal state, soft label $A$ dominates soft label $B$ when

$$g_A \le g_B, s_A \le s_B \quad (9)$$

The first active destination label returned gives the route with the lowest generalized walking cost. If no such label is found, no route exists.

Algorithm 2. Soft rest-sensitive shortest-path search.

```
INPUT  graph G, origin o, destination d, threshold T, penalty λ
Q ← priority queue ordered by generalized cost
A[q] ← active labels at traversal state q
insert Label(o, cost = 0, sinceRest = 0) into Q and A[o]

while Q is not empty do
   L ← extract the label with minimum generalized cost
   if L is inactive then continue
   if L is at d then return reconstructPath(L)

   for each permitted outgoing edge e do
      C ← copy(L)
      for each ordered subsegment x of e do
         ΔE ← max(0, C.sinceRest + x − T)
             − max(0, C.sinceRest − T)
         C.cost ← C.cost + x + λ × ΔE
         C.sinceRest ← C.sinceRest + x
         if x ends at a rest then C.sinceRest ← 0
      if some label in A[state(C)] dominates C then discard C
      else
         mark every label dominated by C inactive
         insert C into A[state(C)] and Q

return NO PATH
```

Appendix A.3 establishes the correctness and termination of Algorithm 2.

#### 2.4.6. Implementation properties

Immediate reversal on the edge just traversed is prohibited to avoid artificial out-and-back rests. To avoid a separate search for every origin and destination, each destination category is evaluated with a bounded reverse multi-source search. All matched destination nodes in a category enter the priority queue with zero initial cost, and the search propagates through the reversed directed pedestrian network. Reversed rest offsets are processed so that each reversed path corresponds to a permitted forward path in the original network. A single search for a category and threshold therefore returns the minimum route to any destination in that category for all origins.

The workflow was implemented in Python 3. OSMnx was used to acquire and prepare OpenStreetMap data; GeoPandas, Shapely and pyogrio handled spatial processing and file input and output; pandas, NumPy and SciPy supported tabular calculations and spatial indexing; and Matplotlib produced the figures.

All data, code and processing scripts required to reproduce the analysis are available in an online repository.

### 2.5. Application and evaluation

#### 2.5.1. Scenario definitions

The primary analysis uses a total walking budget of 1,000 m. At an illustrative walking speed of 1.0 m/s, this corresponds to approximately 16.7 minutes of walking. This is a conservative estimate of walking speed among active older adults (Dommershuijsen et al., 2022). Under the baseline and strict models, the 1,000 m budget applies to physical route length. Under the soft model, it applies to physical distance plus the distance-equivalent penalty for long gaps between rests.

The strict rest-continuity analysis primarily uses maximum walking distances of 100, 250 and 500 m without an opportunity to rest. The 100 m threshold is a stringent precautionary provision standard, consistent with inclusive-mobility guidance for some mobility-impaired pedestrians (UK Department for Transport, 2022). The 500 m threshold is an empirical behavioural reference point after Huber et al. (2026). We also include 250 m as an intermediate scenario.

The soft analysis uses a 500 m resting threshold, with each metre walked beyond it adding half a metre to the modeled walking cost. This penalty rounds the estimated 0.41 m per excess metre derived from Huber et al. (2026) at the study's mean route length; a stronger penalty of one metre is tested in Feudenheim as a robustness check.

Using the notation defined above, the walking budget is $B$ = 1,000 m, and the strict thresholds are $T$ = 100, 250 and 500 m. The soft analysis uses $T$ = 500 m and $\lambda$ = 0.5, 1.0.

#### 2.5.2. Comparative route analysis

For each residential origin, we first identify the baseline or shortest route to a destination in each category and check whether it falls within the walking budget. We then assess whether these same routes meet the selected resting-distance limit, without searching for alternatives. Strict routing searches for the shortest route to a destination in each category that meets both the resting-distance limit and the walking budget. For each residential origin, we then compare the number of accessible categories (zero to five) under baseline routing, shortest-route compliance and strict rerouting. For categories reachable by the ordinary shortest route but not meeting the resting-distance limit, we calculate the additional distance to retain category access as the difference between the shortest compliant alternative and the ordinary shortest route. The compliant alternative may reach a different destination within the same category.

Every category accessible under strict routing is also accessible under soft routing, because routes meeting the resting-distance limit incur no additional cost. The difference between the two measures identifies categories accessible only when longer stretches without rest are allowed at an additional cost. We also measure excess walking burden as the difference between the minimum soft walking cost and the shortest physical distance to a destination in the same category, expressed in distance-equivalent metres. This captures both additional walking and penalties for long stretches without rest. We also calculate the generalized-cost saving achieved by soft rerouting compared with following the ordinary shortest route.

Both additional walking and excess walking burden are weighted by older population. For excess walking burden, we distinguish two averages. The pair-level average combines all baseline-accessible origin–category pairs, weighted by the older population at each origin. The origin-level average first averages burden across baseline-accessible categories at each origin, then weights these averages by older population.

### 2.5.3. Spatial aggregation and reporting

Feudenheim results are mapped at individual residential buildings, while citywide results are aggregated into 500 m hexagons and weighted by estimated older population. Routes may extend beyond these reporting areas, including Mannheim's municipal boundary, to reach destinations and resting opportunities. Hexagons containing fewer than 25 estimated older residents are masked and those without baseline access have no strict-loss value.

## 3. Results

### 3.1. Toy-network demonstration

Figure 1 uses three parallel routes between one origin and destination, with a 100 m maximum distance without rest and a penalty of one equivalent metre for each metre beyond that distance. Route P0 is the 160 m physical shortest path and contains no rest. Route P1 is 190 m long and contains one rest, dividing the journey into stretches of 80 and 110 m. Route P2 is 230 m long and contains two rests, producing stretches of 80, 80 and 70 m.

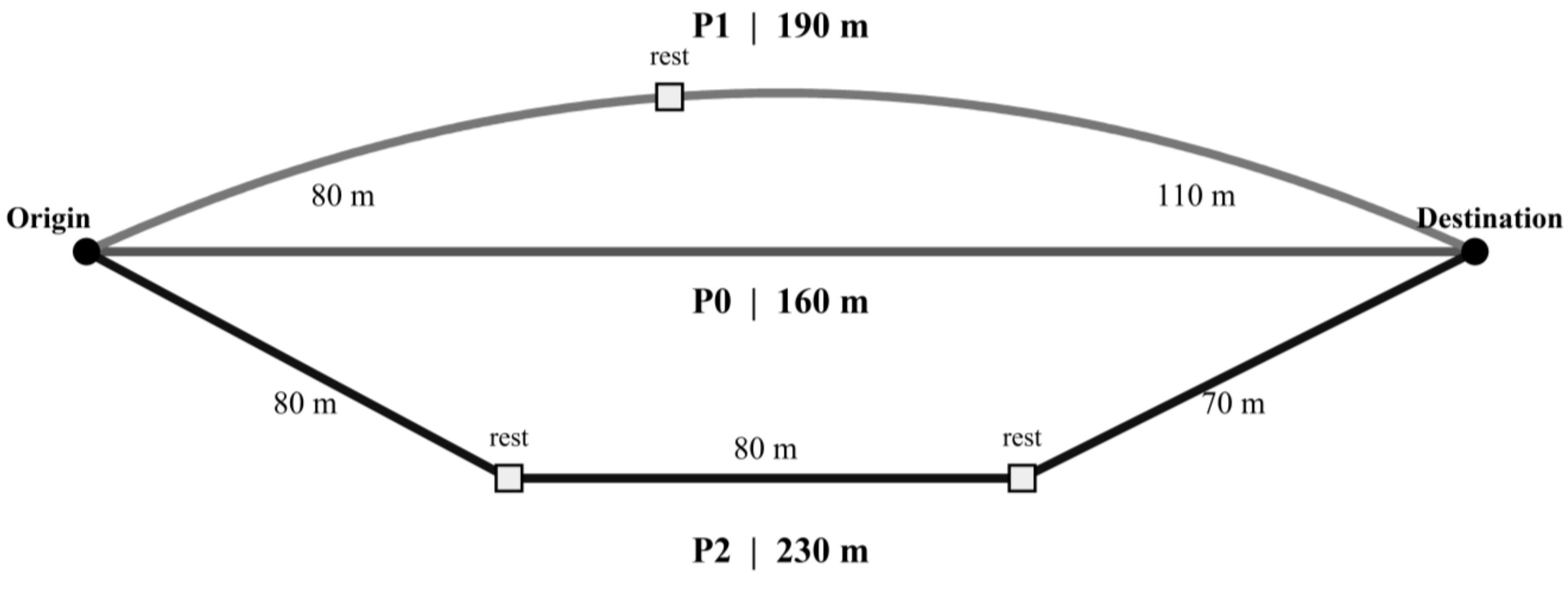

Figure 1. Toy network demonstrating shortest-distance, strict rest-continuity and soft rest-sensitive routing.

The three routing representations answer different questions on the toy network. Shortest-distance routing selects P0. Strict rest-continuity routing rejects P0 and P1 because their longest stretches without rest exceed 100 m and therefore selects P2, the shortest compliant route, adding 70 m of physical walking relative to P0. Under the soft formulation, generalized costs are 220 m for P0, 200 m for P1 and 230 m for P2. It therefore selects P1. The excess burden is 40 equivalent metres: the optimized cost of 200 m remains 40 m above the 160 m physical shortest distance, which would carry no additional penalty if rests were sufficiently spaced.

### 3.2. Analytical coverage

The pedestrian network contained 65,301 network nodes, 179,026 directed edges and 3,935 mapped resting-objects (Table 3). Within Mannheim, 2,898 objects were recorded: 2,182 standalone benches, 643 transit features with seating and 73 picnic tables. Of these, 2,799 (96.6%) matched a pedestrian edge within 20 m; coincident positions reduced them to 2,735 unique routing rest events. The median match distance was 2.9 m (95th percentile: 12.0 m). Mannheim thus contained 18.9 rest events per square kilometer, 0.90 per kilometer of pedestrian network and 43.9 per 1,000 residents aged at least 65 years. Population allocation produced 10,313 weighted network origins with 62,266 older residents. Table 3 details the matched residential locations and everyday destinations used in the routing analysis.

Table 3. Inputs and analytical coverage.

| Input | Source | Available | Used | Role |
|---|---|---|---|---|
| Walking network | OpenStreetMap /OSMnx | 65,301 nodes; 179,026 directed edges | Full support-area graph for routing; 3,040.5 km network in Mannheim for supply/proximity | Route graph |
| Resting opportunities | OpenStreetMap benches, bench-tagged transit features and picnic tables | 2,898 objects within Mannheim | 2,799 matched; 2,735 unique route events | Reset distance since rest |
| Residential origins | OpenStreetMap buildings and 2022 Census grids | 27,589 buildings; 3,872 Census cells | 26,707 buildings; 714 matched residual centroids; 10,313 nodes | Population exposure |
| Community services | | 387 records | 384 matched; 353 nodes | |
| Food shops | | 605 records | 604 matched; 537 nodes | |
| Pharmacies | OpenStreetMap | 125 records | 124 matched; 124 nodes | Everyday destination |
| Primary healthcare | | 542 records | 536 matched; 389 nodes | |
| Public transport | | 2,173 records | 1,591 matched; 1,591 nodes | |

### 3.3. Neighbourhood-scale rest provision and route accessibility in Feudenheim

#### 3.3.1. Seating supply and proximity

Feudenheim had 209 mapped resting opportunities, of which 207 matched the pedestrian network, yielding 205 unique rest positions after duplicates were removed. The reporting area contains 155.2 km of physical pedestrian network and approximately 3,565 estimated older residents. This corresponds to 33.7 modeled rest positions per km$^2$, 1.32 per km of physical pedestrian network, and 57.5 per 1,000 older residents.

The local pattern was not spatially uniform (Figure 2). The median walking-network distance to the nearest rest was 194 m across the pedestrian network, with a 90th percentile of 499 m and a 95th percentile of 704 m. Respectively, 25.3%, 62.8% and 89.9% of pedestrian-network length lay within 100, 250 and 500 m of a mapped rest. Distances from homes were more favourable at the wider thresholds: the older-population-weighted median distance from home to the nearest rest was 197 m, and 16.2%, 68.6% and 99.5% of the older population lived within 100, 250 and 500 m, respectively.

### 3.3.2. Shortest-route threshold compliance

Under ordinary shortest-distance routing, 99.4% of older residents could reach at least one destination category within 1,000 m, 89.7% could reach all five, and mean coverage was 4.86 categories. Applying resting-distance limits of 100, 250 and 500 m to these same routes reduced mean compliant coverage to 0.47, 1.83 and 3.78 categories, respectively. The corresponding shares reaching all five categories were 0%, 4.3% and 33.5% (Figure 3). The effect also differed strongly by destination type (Table 4). With a maximum of 250 m between rests, shortest-route compliance ranged from 12.3% for pharmacy access to 54.5% for public transport. With a 500 m maximum, public transport was already largely compliant on shortest routes (98.0%), whereas pharmacy access remained compliant for only 40.0% of older residents. Thus, local seating provision and even a nearby rest at home do not establish that a useful rest sequence exists on the routes required for different everyday activities.

### 3.3.3. Strict continuity recovery and additional walking

Strict rerouting increased mean coverage from 0.47 to 0.51 categories at the 100 m resting-distance limit, from 1.83 to 2.28 at 250 m, and from 3.78 to 4.54 at 500 m (Table 4). The corresponding shares reaching all five categories increased from 0% to 0%, 4.3% to 6.8%, and 33.5% to 61.4%, respectively (Figure 4). At the 250 m resting-distance limit, strict rerouting produced the largest gains in primary healthcare (+17.9 percentage points), food shops (+10.9) and public transport (+9.5) (Table 4). At 500 m, the largest gains were in community services (+29.6 points) and pharmacies (+22.4 points). Pharmacy access nevertheless remained lowest at 62.3%, compared with 89.7% under ordinary shortest-distance routing. Strict access to food shops, primary healthcare and public transport reached 97.6%, 98.6% and 99.1%, respectively.

Of the categories accessible under ordinary shortest-distance routing, 1.0%, 9.2% and 15.6% could meet the respective 100, 250 and 500 m resting limits only by taking an alternative route. Across recovered cases, mean additional walking was 104 m (32.3%) at the 250 m limit and 88 m (14.0%) at 500 m; corresponding medians were 65 m (19.5%) and 82 m (11.1%). At 250 m, food shops and pharmacy access required the greatest mean additional distances (160 m and 131 m, respectively). At 500 m, strict rerouting recovered at least one category for 50.3% of older residents, while 30.5% still lost access to at least one baseline-accessible category.

### 3.3.4. Soft rest-sensitive accessibility and excess walking burden

Soft routing yielded mean coverage of 4.71 categories, compared with 4.54 under strict routing and 4.86 under ordinary shortest-distance routing. The corresponding shares of older residents reaching all five categories were 76.2%, 61.4% and 89.7%. Compared with strict routing, soft routing provided access to at least one additional category for 16.6% of older residents, with an average gain of 0.17 categories per origin. It increased pharmacy access by 13.9 percentage points with gains in other categories ranging from 0 to 1.7 points (Table 4).

The population-weighted mean excess walking burden was 12.5 distance-equivalent metres across baseline-accessible categories, reaching 42.2 m for pharmacies and 14.5 m for community services. Among older residents at origins with excess burden above 25 m, 78.3% gained no additional category under soft routing. Soft rerouting reduced generalized walking cost in 10.5% of soft-accessible origin–category cases. In these cases, the mean saving was 74.2 distance-equivalent metres despite 27.1 m of additional physical walking. In 13.9% of soft-accessible cases, the selected route still included a stretch longer than 500 m without rest.

Doubling the soft penalty to one added equivalent metre per metre beyond the 500 m no-rest distance did not change this interpretation. Mean coverage declined slightly from 4.71 to 4.69 categories, the share reaching all five categories from 76.2% to 74.4%, and the share gaining at least one category from 16.6% to 14.3%. Pharmacy remained the dominant gain under soft rather than strict routing.

Table 4. Category-specific access in Feudenheim under ordinary shortest-path, strict-continuity and soft rest-sensitive routing, using a 1,000 m total walking budget.

| Destination category | Baseline shortest (%) | Shortest compliant (%) 100 m | Strict (%) 100 m | Shortest compliant (%) 250 m | Strict (%) 250 m | Shortest compliant (%) 500 m | Strict (%) 500 m | Soft (%) 500 m, 0.5 penalty |
|---|---|---|---|---|---|---|---|---|
| Community services | 98.8 | 3.9 | 5.0 | 23.0 | 27.2 | 67.0 | 96.5 | 98.2 |
| Food shops | 98.8 | 9.1 | 10.7 | 39.5 | 50.4 | 84.0 | 97.6 | 98.7 |
| Pharmacy | 89.7 | 3.0 | 3.0 | 12.3 | 14.3 | 40.0 | 62.3 | 76.2 |
| Primary healthcare | 99.3 | 15.3 | 15.4 | 53.7 | 71.6 | 89.3 | 98.6 | 99.2 |
| Public transport | 99.1 | 15.2 | 17.1 | 54.5 | 64.0 | 98.0 | 99.1 | 99.1 |
| Mean number of categories | 4.86 | 0.47 | 0.51 | 1.83 | 2.28 | 3.78 | 4.54 | 4.71 |
| Share reaching all five categories (%) | 89.7 | 0.0 | 0.0 | 4.3 | 6.8 | 33.5 | 61.4 | 76.2 |

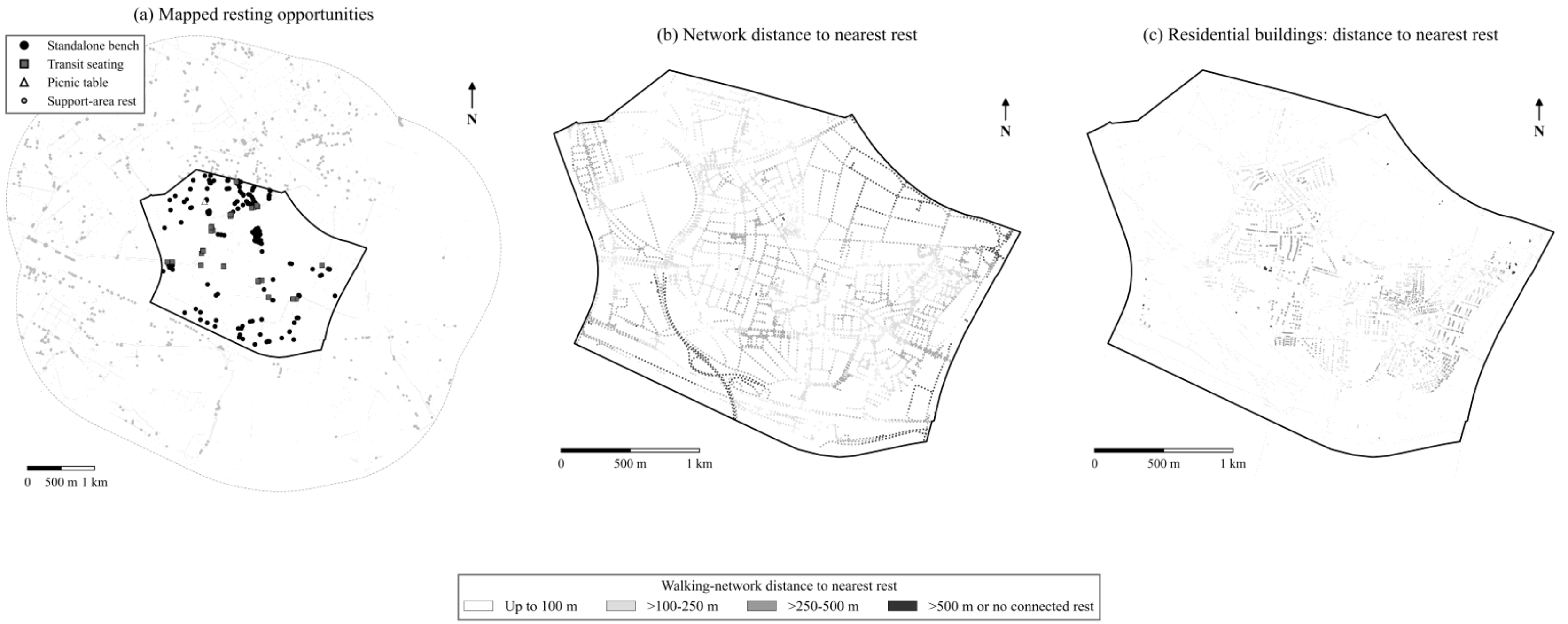

Figure 2. Seating supply and proximity in Feudenheim. Panel (a) shows mapped resting opportunities attached to the pedestrian network; pale symbols outside the reporting boundary are retained in the routing support area but are not counted as Feudenheim provision. Panel (b) classifies 25 m pedestrian-network samples by walking-network distance to the nearest rest. Panel (c) classifies residential building footprints by the distance from their matched origin to the nearest rest. Distances above 500 m and locations without a connected rest are combined.

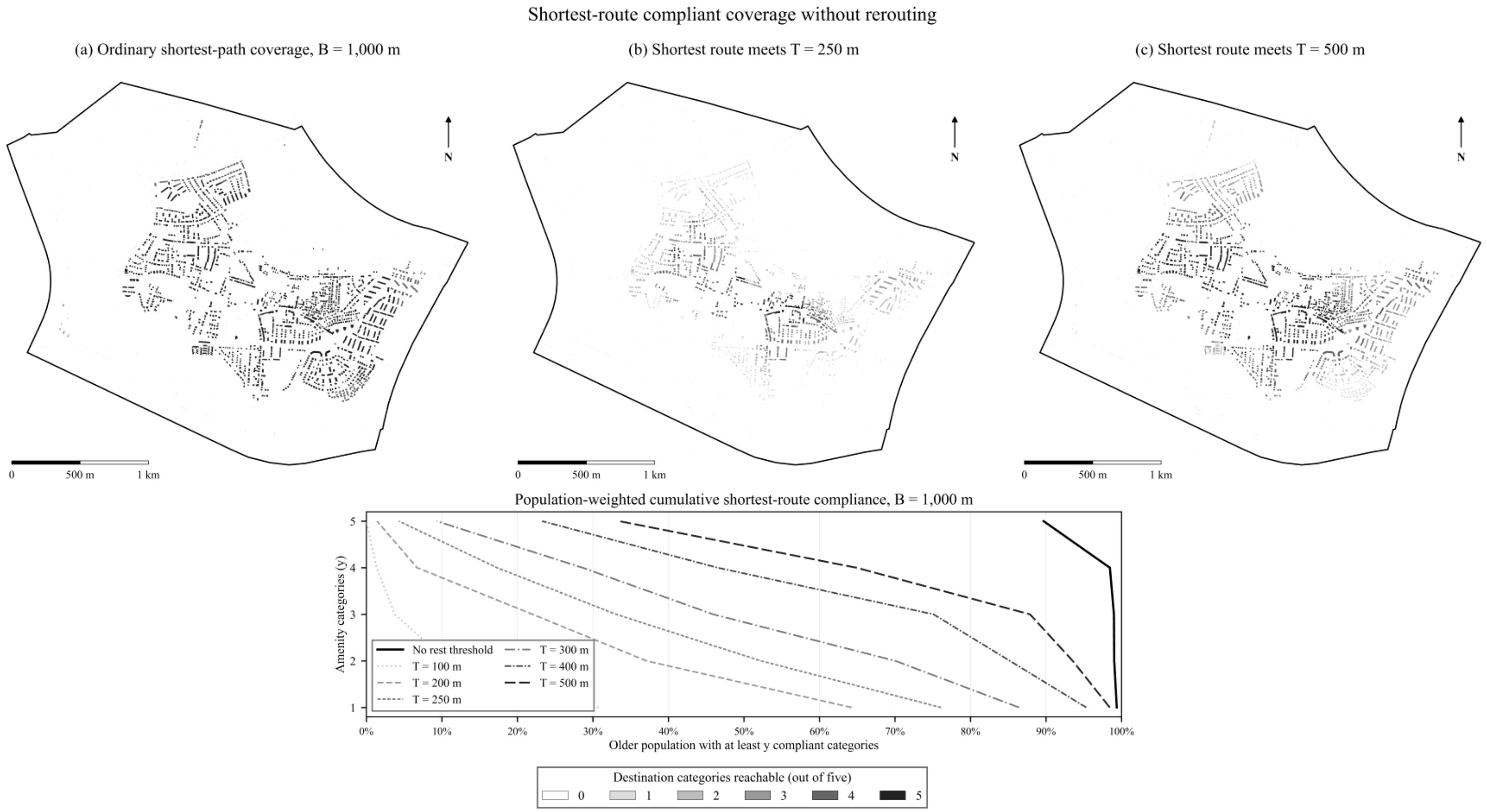

Figure 3. Ordinary shortest-route coverage before strict rerouting in Feudenheim, with a 1,000 m total walking budget. Panel (a) maps the number of destination categories reachable by an ordinary shortest route. Panels (b) and (c) show how many of those shortest routes also meet maximum distances of 250 m and 500 m between rests, respectively. The lower panel shows the older-population-weighted cumulative distribution of compliant category coverage across multiple resting thresholds.

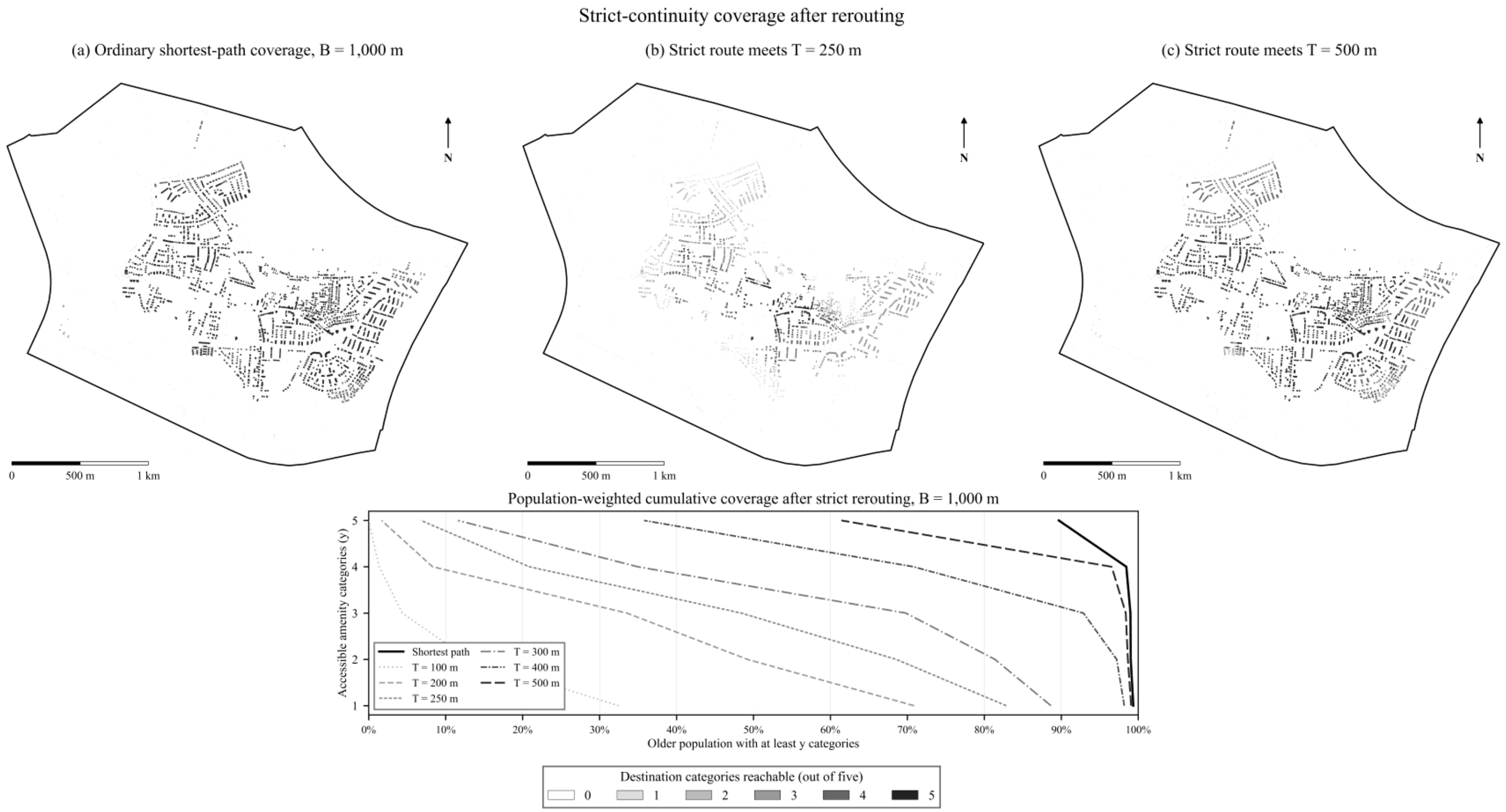

Figure 4. Strict-continuity route coverage after rerouting in Feudenheim, with a 1,000 m total walking budget. Panel (a) maps ordinary shortest-path coverage. Panels (b) and (c) show the number of destination categories reachable by at least one route on which no stretch without an opportunity to rest exceeds 250 m or 500 m, respectively. The lower panel shows the older-population-weighted cumulative distribution of strict category coverage across multiple tested thresholds.

### 3.4. Citywide heterogeneity in rest-sensitive accessibility

#### 3.4.1. Seating supply and proximity

Resting opportunities were concentrated in Mannheim's built-up areas and were sparser in peripheral agricultural, forest and industrial areas (Figure 5a), contributing to the long upper tail of nearest-rest distances (Table 5). However, gaps in provision also occurred within built-up areas (Figure 5b). Older residents generally lived closer to resting opportunities than the pedestrian network as a whole, reflecting their concentration in built-up areas (Figure 5c; Table 5).

#### 3.4.2. Citywide route-level accessibility, continuity deficit and excess burden

The 274 retained hexagons covered 98.6% of Mannheim's estimated older population. Under ordinary shortest-distance routing, mean coverage was 4.71 of five destination categories, and 84.3% of older residents could reach all five within 1,000 m. Access was lowest for pharmacies (87.1%) and highest for public transport (98.3%) (Figure 6; Table 6).

Under strict routing, 7.2% of baseline-accessible origin–category pairs lost access, affecting 20.2% of older residents. Across hexagons, mean strict loss had a median of 0.10 categories and a 90th percentile of 1.65. Losses were greatest for pharmacies and lowest for public transport. Under soft routing, the mean excess walking burden was 10.9 distance-equivalent metres across residential origins. Across hexagons, the median burden was 7.3 m (90th percentile: 45.9 m), with pharmacies showing the highest category-specific mean (19.8 m). Soft routing retained 5.1 percentage points more category access than strict routing.

Table 5. Walking distances to the nearest rest across the pedestrian network and from homes.

| Metric | Physical network | Older-population origins |
|---|---|---|
| Connected to a mapped rest | 98.3% | 98.6% |
| Median distance | 246.7 m | 181.9 m |
| 90th-percentile distance | 944.2 m | 374.2 m |
| 95th-percentile distance | 1,308.8 m | 460.2 m |
| Within 100 m | 18.8% | 20.7% |
| Within 250 m | 49.7% | 70.4% |
| Within 500 m | 74.3% | 95.2% |

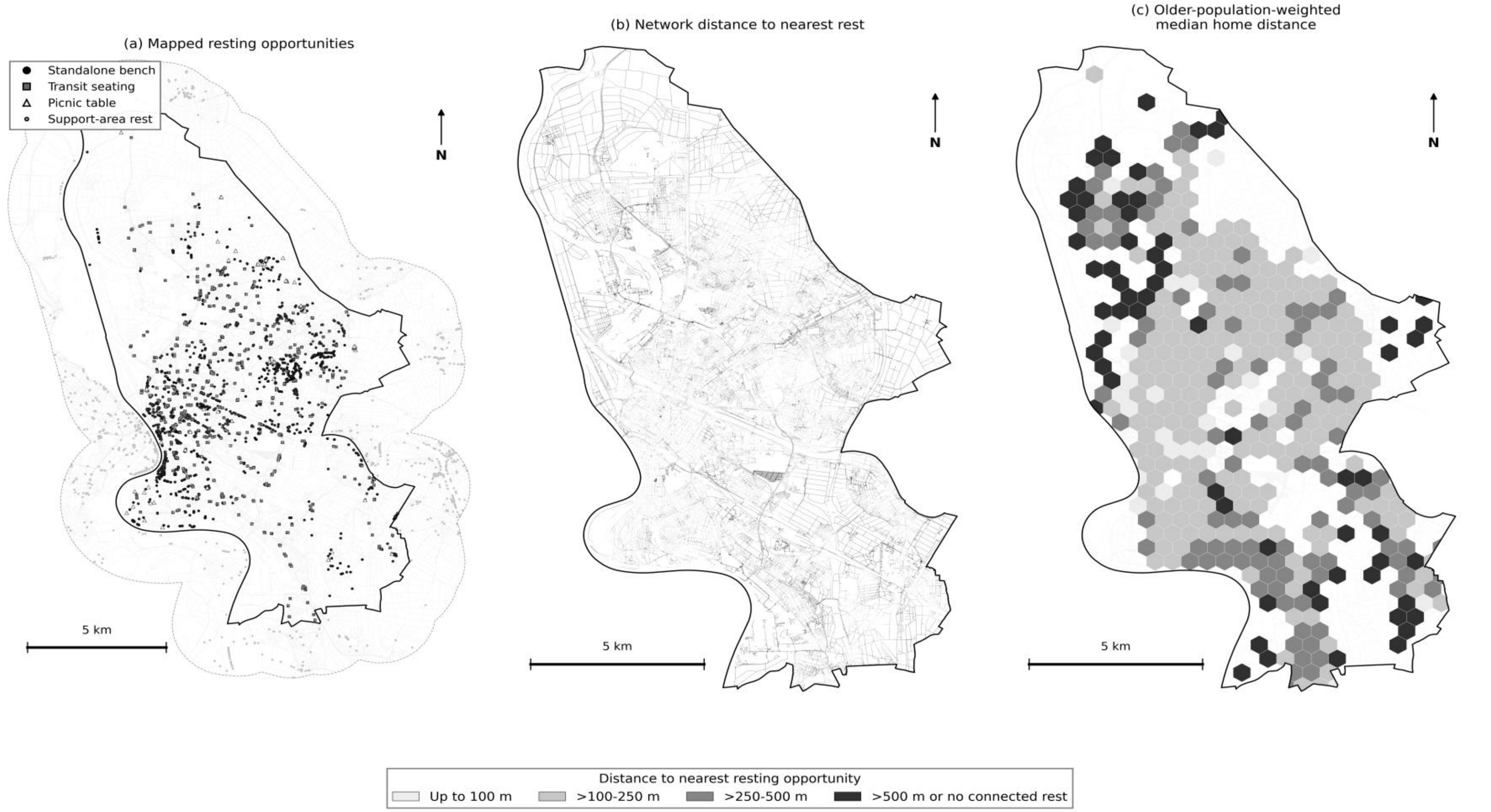

Figure 5. Seating supply and proximity in Mannheim. Panel (a) shows matched resting opportunities by source type. Panel (b) classifies 25 m pedestrian-network samples by shortest network distance to the nearest rest. Panel (c) maps the median distance from homes to the nearest rest in 500 m hexagons, weighted by older population.

Table 6. Citywide category-specific route access among Mannheim older residents, with a 1,000 m journey budget, a 500 m reference distance between rests, and an added soft-routing cost of half an equivalent metre per metre beyond that distance.

| Destination category | Baseline shortest (%) | Shortest compliant (%) | Strict (%) | Soft (%) | Soft minus strict (pp) | Mean excess burden (m) |
|---|---|---|---|---|---|---|
| Food shops | 96.7 | 80.7 | 90.7 | 95.6 | +4.9 | 9.2 |
| Pharmacies | 87.1 | 56.6 | 75.3 | 83.4 | +8.1 | 19.8 |
| Primary healthcare | 94.5 | 77.9 | 87.2 | 92.4 | +5.1 | 11.1 |
| Public transport | 98.3 | 95.8 | 97.2 | 98.3 | +1.1 | 1.0 |
| Community service | 93.8 | 74.2 | 86.1 | 92.2 | +6.0 | 11.4 |
| All category opportunities | 94.1 | 77.0 | 87.3 | 92.4 | +5.1 | 10.3 |

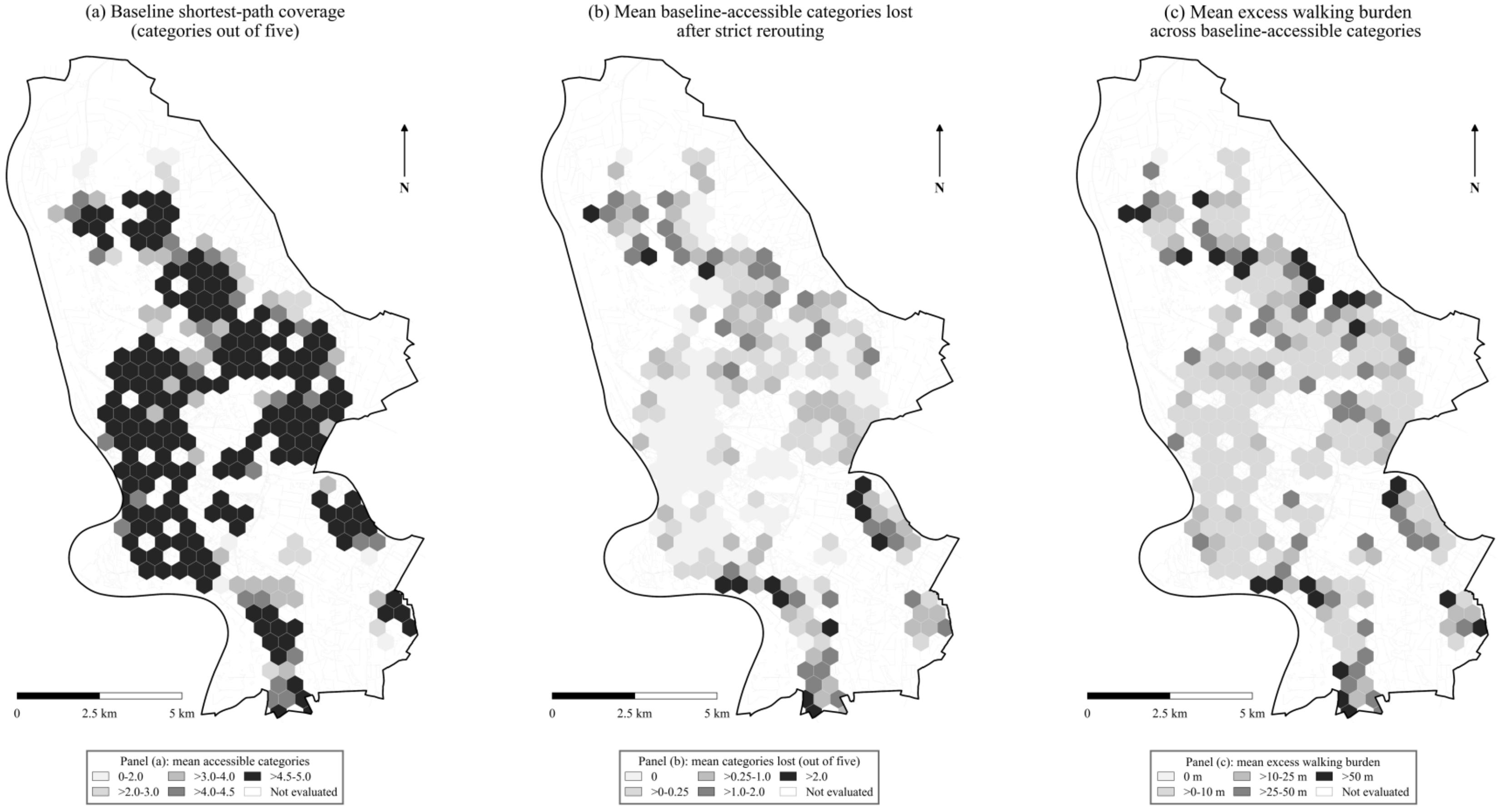

Figure 6. Citywide accessibility and excess walking burden in Mannheim, using a 1,000 m journey budget and a 500 m resting threshold. Population-weighted 500 m hexagons show (a) mean category coverage under ordinary shortest-distance routing, (b) mean categories lost after strict rerouting, and (c) mean excess walking burden under soft routing. Soft routing adds half a distance-equivalent metre per metre beyond the resting threshold.

## 4. Discussion

The findings show that nearby benches and physically reachable destinations do not necessarily provide journeys with well-spaced resting opportunities. Feudenheim illustrates this mismatch at individual residential origins, while the Mannheim-wide analysis reveals different spatial patterns of limited destination access. Distinguishing these conditions is more informative for planning than combining them into a single accessibility index.

### 4.1. From diagnosis to planning action

A planning department could use the framework outputs to identify residential areas and routes for investigation, then simulate adding or removing benches to assess changes in access, detours and excess walking burden at buildings or across a neighbourhood. The same calculations could assess the addition, removal or relocation of amenities, or evaluate accessibility from proposed residential developments. Population weights would indicate how many older residents could benefit.

Computationally selecting bench locations that maximize access or minimize excess walking burden under a seating budget requires an additional facility-location optimization model. This requires defining candidate sites, objectives and constraints, and accounting for benefits that depend on combinations of benches and the routes they enable. Developing this optimization is a subject for future work.

### 4.2. Mannheim: a walkable city for older adults

Although services and seating are concentrated in Mannheim's inhabited areas, proximity to a rest does not guarantee adequately spaced opportunities along complete journeys. Under the 500 m resting-distance limit, 20.2% of older residents lost access to at least one otherwise reachable category. Allowing longer stretches at additional modeled cost reduced this share to 7.1%. In Feudenheim, pharmacies showed the greatest loss under strict routing and the largest recovery under soft routing, while public transport remained comparatively accessible. This may reflect the smaller number of pharmacies and alternative routes, compared with the wider distribution of public-transport stops, some of which provide seating. The citywide results distinguish areas with limited destination access, which benches alone cannot resolve, from areas where insufficiently spaced rests restrict access or increase walking burden. The latter conditions can help identify where targeted seating interventions may be most useful.

### 4.3. From accessibility assessment to personal navigation

Beyond infrastructure planning, rest-sensitive routing could also help pedestrians choose routes suited to their walking capabilities rather than assuming that the shortest route is always appropriate (Fulman et al., 2025a).

A navigation system could offer three alternatives: the ordinary shortest route, the shortest route meeting a user's resting-distance limit, and the least-burdensome route when exceeding that limit is allowed at additional cost. Presenting their physical distances and resting opportunities would help travellers compare the alternatives. Resting-distance thresholds could be selected by the traveller. In future, such applications could use large language models to interpret resting needs expressed in natural language, although reliable handling of individual preferences remains challenging (Fulman et al., 2025b; Song et al., 2026). Personalization would require further behavioural validation, as older adults vary in how far they can or prefer to walk without rest.

Rest-sensitive routing could also form one component of a broader pedestrian navigation system. Distance could be considered alongside slope, surface conditions, crossings, traffic exposure, shade and greenery, building on existing green and shaded routing approaches (Novack et al., 2018; Ludwig et al., 2021; Helle et al., 2023; Feng et al., 2024). The underlying reset logic may also be transferable to other journeys involving recurring service needs, such as access to restrooms, vehicle charging or scheduled driver rests, as well as tourism routes where reasonably spaced opportunities to sit improve a long walk between landmarks.

### 4.4. Limitations

First, the analysis depends on the completeness and accuracy of OpenStreetMap's pedestrian network, resting opportunities and destinations. Missing or incorrectly positioned features can affect route access despite the network-matching tolerances; checks in Appendix B support the data's suitability but do not eliminate uncertainty. Missing building levels also introduce uncertainty in population allocation. Assuming all mapped seats are usable may overestimate strict accessibility and underestimate soft walking burden. More detailed inventories could distinguish benches by features and condition, while real-time information could identify temporary unavailability due to occupancy or wetness. Both could be represented by updating the set of usable resting opportunities without changing the routing algorithms. In personal navigation, travellers could similarly exclude benches they prefer not to use. Explicitly modelling a decision to stop, including resting time, would require extending the model to balance stopping against continuing.

Second, residential origins and destinations are matched to network nodes within 75 m. The connecting distances are excluded from route length, resting-distance calculations and generalized cost. This may affect accessibility classifications near the selected limits, particularly in the strict 100 m scenario.

Third, the framework isolates the role of resting opportunities in destination accessibility rather than providing a complete measure of older-adult walkability. Routes are evaluated on a planar pedestrian network, so differences in walking effort due to slopes, elevation changes and vertical infrastructure such as footbridges and underpasses are not represented. Recent studies have incorporated three-dimensional pedestrian networks and terrain-related walking impedance into accessibility assessments for older adults (Liu et al., 2026; Ng et al., 2024). Although Mannheim is relatively flat, incorporating these features would extend the framework to more topographically complex settings, both for accessibility assessment and personalized navigation.

Finally, the model excludes surface conditions, crossings, shade, greenery, weather, destination quality and opening hours, bench occupancy and resting time. Accessibility is measured by the ability to reach at least one destination in each category, rather than accounting for the frequency or combination of journeys people actually make. Walking distance limits and the additional cost of walking beyond them without an opportunity to rest are also specified for each scenario rather than reflecting individual differences in walking capabilities and preferences. The results therefore describe potential accessibility and rest-related burden under the selected scenarios, rather than observed travel behaviour.

Declaration of Generative AI: Use During the preparation of this manuscript, the authors used ChatGPT (GPT-5.6 Sol, OpenAI) to assist with language editing, improving clarity, and refining the wording of selected passages. The authors reviewed and edited all AI-assisted content and take full responsibility for the final manuscript.

Appendix A. Correctness and optimality

A.1. Common assumptions and notation

The arguments apply to a finite directed pedestrian graph in which every edge has strictly positive length and every permitted transition is available to all labels at the same traversal state. Rest offsets may be viewed as vertices in an expanded graph, so the number of nodes, edges and rest events is finite. The threshold satisfies $T \geq 0$, the soft penalty satisfies $\lambda \geq 0$, and comparisons are stated in exact arithmetic. The implementation may use a small numerical tolerance without changing the logical inequalities. A feasible route also obeys the stated turn restriction, including the prohibition on immediate reversal.

Labels are compared only when they occupy the same traversal state $q$ and therefore have identical permitted continuations. A rest resets $s$ to zero after the distance and, for the soft model, the cost of reaching that rest have been charged. The shortest-distance baseline is ordinary Dijkstra search and requires no additional argument; the results below concern the two rest-sensitive formulations.

A.2. Strict rest-continuity algorithm

Lemma A.1. Monotonicity of strict continuation feasibility

If two labels at the same traversal state satisfy $s_A \leq s_B$, every fixed continuation that is feasible from $B$ is also feasible from $A$.

Proof. Until the first rest on the continuation, the distance-since-rest values differ only in their initial values, so $s_A + y \leq s_B + y$ for every traversed distance $y$. If the continuation never exceeds $T$ from $B$, it cannot exceed $T$ from $A$. At the first rest, both distances since rest reset to zero and subsequently evolve identically. The same conclusion holds for the remainder of the continuation. QED.

Lemma A.2. Safety of strict dominance

Suppose strict labels $A$ and $B$ occupy the same traversal state and satisfy $d_A \leq d_B$ and $s_A \leq s_B$. Discarding $B$ cannot remove the unique prefix of a shortest compliant route.

Proof. Consider any continuation $R$ that completes $B$ to the destination without exceeding $T$. By Lemma $A$.1, $R$ is also feasible from $A$. Its physical length is independent of the prefix, and $d_A \leq d_B$, so completing $A$ along $R$ produces a route no longer than the corresponding completion of $B$. Thus $B$ cannot lead to a strictly better compliant route and may be discarded. QED.

Lemma A.3. Strict label-setting order

Every feasible extension of a strict label increases its accumulated distance.

Proof. Every nonempty extension contains strictly positive physical length, which is added to $d$. A descendant therefore has greater distance than its parent. Because the queue is ordered by $d$, no unprocessed descendant can have a value below the active queue minimum. QED.

Theorem A.1. Optimality of the strict route

If Algorithm 1 returns a destination label, its reconstructed route has minimum physical length among all routes satisfying $G(P) \leq T$. If the algorithm returns NO COMPLIANT PATH, no such route exists.

Proof. Let the first active destination label removed from the queue have distance $d^*$. Assume that a compliant route $P$ has length less than $d^*$. Follow the labels corresponding to successive prefixes of $P$. Whenever a prefix label is discarded, Lemma A.2 supplies a dominating label that can follow the same remaining route without losing feasibility or increasing its final length. Repeating this replacement leaves an active nondominated representative of an unprocessed prefix with distance less than $d^*$. Lemma A.3 requires that representative to be processed before the destination label, a contradiction. The returned route is therefore optimal. The same prefix argument shows that if any compliant route existed, some active representative of it would reach the destination; hence an empty queue implies that none exists. QED.

Theorem A.2. Termination of the strict search

Under the assumptions in Appendix A.1, Algorithm 1 processes finitely many labels and terminates.

Proof. Between two rest events, a nondominated prefix cannot repeat a traversal state: removing the resulting bench-free cycle reduces both $d$ and $s$ while preserving all later feasibility. At a rest event, $s$ resets to zero. Returning through a positive-length cycle to the same traversal state immediately after a rest therefore yields zero distance since rest with greater $d$ and is dominated; repeating such a cycle cannot create further nondominated labels. Because the expanded graph and its rest events are finite, only finitely many nonrepeating traversal-state and rest-event sequences remain. The algorithm can accept only finitely many nondominated labels before it reaches a destination or empties the queue. QED.

A.3. Soft rest-sensitive algorithm

Lemma A.4. Monotonicity of soft continuation cost

For a fixed continuation, its additional generalized cost is nondecreasing in the initial distance since rest $s$.

Proof. For a bench-free subsegment of length $x$, $\Delta E(s, x; T)$ is zero when $s \leq T - x$, equals $s + x - T$ when $T - x < s < T$, and equals $x$ when $s \geq T$. It is therefore nondecreasing in $s$. The physical-distance term $x$ does not depend on $s$ and $\lambda \geq 0$, so the complete subsegment cost is also nondecreasing. If a continuation contains a rest, only the portion before the first rest depends on the initial distance since rest; this distance then resets to zero. The result holds for every continuation. QED.

Lemma A.5. Safety of soft dominance

Suppose soft labels $A$ and $B$ occupy the same traversal state and satisfy $g_A \leq g_B$ and $s_A \leq s_B$. Discarding $B$ cannot remove the unique prefix of a minimum-generalized-cost route.

Proof. Let $\Delta(R, s)$ be the additional generalized cost of any fixed continuation $R$ entered with distance since rest $s$. Lemma $A$.4 gives $\Delta(R, s_A) \leq \Delta(R, s_B)$. Combining this with $g_A \leq g_B$ gives $g_A + \Delta(R, s_A) \leq$ $g_B + \Delta(R, s_B)$. Thus no completion of $B$ can be cheaper than the corresponding completion of $A$, and $B$ may be discarded. QED.

Lemma A.6. Soft label-setting order

Every extension of a soft label increases its accumulated generalized cost.

Proof. Every extension adds strictly positive physical distance and the non-negative term $\lambda \Delta E$. A descendant therefore has greater $g$ than its parent. Because the queue is ordered by $g$, no unprocessed descendant can have a value below the active queue minimum. QED.

Theorem A.3. Optimality of the soft rest-sensitive route

When Algorithm 2 removes its first active destination label from the queue, the reconstructed route minimizes $C_{T,\lambda}(P)$ over all permitted routes between the origin and destination.

Proof. Let that destination label have cost $g^*$. Assume for contradiction that a route $P$ has generalized cost below $g^*$. Follow the labels associated with successive prefixes of $P$. Whenever one is discarded, Lemma A.5 allows it to be replaced by a dominating label whose completion along the same remaining route is no more costly. An active nondominated representative of an unprocessed prefix must therefore exist with accumulated cost below $g^*$. By Lemma A.6, it must be removed before the destination label, contradicting the definition of $g^*$ as the active queue minimum. No cheaper route exists. QED.

Theorem A.4. Termination of the soft search

Under the assumptions in Appendix A.1, Algorithm 2 processes finitely many labels and terminates.

Proof. Between two rest events, repeating a traversal state creates a bench-free cycle. Removing that cycle reduces both $g$ and $s$, so the cycled label is dominated. Returning to the same traversal state immediately after a rest produces $s = 0$ with a strictly greater $g$ and is likewise dominated; repeating the reset cycle cannot generate a new nondominated condition. The finite expanded graph therefore permits only finitely many nonrepeating traversal-state and rest-event sequences. Only finitely many nondominated labels can be accepted, after which the algorithm reaches a destination or empties the queue. QED.

## Appendix B. Input data validation

### B.1. Pedestrian network

To assess the suitability of OpenStreetMap for pedestrian routing, we compared its walking network with official ATKIS Basis-DLM data from the Baden-Württemberg mapping authority (LGL, 2026). Using a 10 m spatial tolerance, we found that 97.6% of official street-axis length, 92.6% of footpath and shared pedestrian–cycle path length, and 92.9% of other path and track length in Mannheim were represented near the OpenStreetMap network. Corresponding agreement in Feudenheim ranged from 93.2% to 94.8%. A complementary comparison, restricted to comparable OpenStreetMap features, found that 81.3% of OpenStreetMap network length in Mannheim (86.5% in Feudenheim) lay within 10 m of ATKIS transport linework. Visual inspection of discrepancies indicated additional fine-scale pedestrian connections in OpenStreetMap, particularly in residential areas. This is consistent with the different mapping objectives: ATKIS selects paths according to their topographic significance, whereas OpenStreetMap also represents smaller connections relevant to pedestrian routing. Overall, the high correspondence with official data, combined with OpenStreetMap's additional pedestrian detail, supports its selection for this study.

### B.2. Resting opportunities

A targeted imagery-based check assessed resting opportunities in parts of central Mannheim, including Quadrate, Neckarstadt, and surrounding neighbourhoods, using randomly selected Mapillary street-level photographs (2026). First, 50 bench locations retained following manual inspection of Mapillary bench detections were compared with the original OpenStreetMap seating inventory. Of these, 45 (90%) had a mapped resting opportunity within a straight-line distance of 20 m. Second, 50 OpenStreetMap seating locations in the same area were examined in random order. Seating was visible at 47 locations (94%) but not at the remaining three (6%). The photographs were taken between 2015 and 2026, meaning that some may predate the installation of benches.

### B.3. Sensitivity to missing building levels

Usable floor counts were recorded for 72.4% of residential buildings in Mannheim and 90.1% in Feudenheim. To assess sensitivity to assigning one level where this information was missing, we retained recorded values and replaced missing values with the observed Mannheim-wide median for each OpenStreetMap building type: three levels for apartments, two for houses, generic residential buildings and terraces, and one for detached houses. We repeated the population allocation within the same 100 m Census cells, retaining the original network matching and population rescaling, and reweighted the existing route results at a 1,000 m walking budget, a 500 m resting threshold and a penalty of 0.5. In Feudenheim, the shares reaching all five destination categories changed from 89.68% to 89.71% under baseline routing, from 61.43% to 61.67% under strict routing, and from 76.24% to 76.21% under soft routing when missing building levels were assigned building-type medians instead of one level. Across the same 274 Mannheim reporting hexagons, median excess walking burden remained 7.3 m, while its 90th percentile increased from 45.9 to 47.4 m. The largest absolute change in an individual hexagon was 3.2 m. Burden rankings remained closely aligned (Spearman correlation 0.99), with 27 of the 28 highest-burden hexagons retained in the highest decile. This alternative allocation did not change the qualitative interpretation of the selected neighbourhood and citywide applications.